# Sequential Information Elicitation in Multi-Agent Systems

**Rann Smorodinsky and Moshe Tennenholtz**
Faculty of Industrial Engineering and Management
Technion – Israel Institute of Technology
Haifa 32000, Israel

## Abstract

We introduce the study of sequential information elicitation in strategic multi-agent systems. In an information elicitation setup a center attempts to compute the value of a function based on private information (a-k-a secrets) accessible to a set of agents. We consider the classical multi-party computation setup where each agent is interested in knowing the result of the function. However, in our setting each agent is strategic, and since acquiring information is costly, an agent may be tempted not spending the efforts of obtaining the information, free-riding on other agents' computations. A mechanism which elicits agents' secrets and performs the desired computation defines a game. A mechanism is 'appropriate' if there exists an equilibrium in which it is able to elicit (sufficiently many) agents' secrets and perform the computation, for all possible secret vectors. We characterize a general efficient procedure for determining an appropriate mechanism, if such mechanism exists. Moreover, we also address the existence problem, providing a polynomial algorithm for verifying the existence of an appropriate mechanism.

## 1 Introduction

Information elicitation deals with the extraction of information from agents. In a multi-agent setup a distinguished agent (a-k-a the center) may be interested in extracting information from a set of agents, each of which has its own private information. In general, it is interested in computing the result of a given function (e.g. the majority function) when applied to the agents' private information (e.g. bit values representing the alternatives they support). A most general variant of this problem is the multi-party computation problem. In that setting the agents themselves are also interested in learning the value of the given function, when applied to the agents' private information (sometime referred to as the agents' secrets). This general setting is central to computer science and reasoning about uncertainty. For example, it captures basic problems in computation based on a distributed data base, in computing a population statistic such as an order statistic, in computing a clearing price in (say) a two-sided auction, and in voting. We address information elicitation in this setting from a game theoretic perspective and introduce a 'multi party computation game'. We assume that the agents are selfish and are driven by utility maximization considerations. On the one hand, they would like to receive the value of the multi party computation, but on the other hand they may not want to contribute to this computation, but rather free-ride on other agents' efforts.

The particular (although very general) setting we study is the following.[1] Each agent has a secret, which is accessible only to him. However, access to that secret is costly and agents may choose to access it, or not.[2] Thus, accessing one's own private information becomes a strategic question. This approach generates a natural tension between the socially (and even privately) optimal action, which is to compute the joint

---

[1] The study presented in this paper shares its motivation, and some of the presentation with a complementary line of research presented in another paper we submitted to a game-theory forum (Smorodinsky & Tennenholtz 2003). The current paper however concentrates, in difference to that other work, on the major problem of finding criteria for the existence of appropriate information elicitation mechanisms, as well as on the computational complexity of verifying the existence of such mechanisms.

[2] One can think of computational cost, cost of accessing a data base, or even a learning cost.

function correctly, and agents' incentive to free-ride. In order to overcome this tension (and enable the elicitation of the desired information) one may need to design an elicitation mechanism to prevent (some or all) agents from free-riding, elicit agents' secrets and execute the desired computation. To better understand the basic tension in the multi-party computation game consider the following:

**Example 1** *Assume there are* 11 *agents, each having a private value which equals either 0 or 1 with equal probabilities and which are independent. The agents would like to compute the simple majority function. Consider a situation where each agent must pay* 0.4 *for accessing his secret, but values the correct result of the computation at* 1. *A simple elicitation mechanism asks the agents, in some arbitrary order, for their secrets. The mechanism will halt when the value of the majority is already determined. The strategy tuple where all agents choose to access their secrets, if/when approached, and report them truthfully is not an equilibrium. To see this consider the perspective of agent* 1, *the first agent to be approached. Assuming all agents will reply (truthfully, or not) if/when approached, then agent* 1 *can alter the outcome of the majority function only if the other* 10 *replies split evenly between* 0 *and* 1, *which has a probability of* $\frac{10!}{5!5!}0.5^{10} \approx \frac{1}{4}$. *Therefore, by accessing his own secret an agent gains, at most (assuming all others access their secrets),* $1 - 0.4 = 0.6$. *However, by guessing, and assuming all other agents compute, he will gain* $0.25 \times 0.5 + 0.75 \times 1 = 0.875$ *(where* 0.25 *is the probability of being pivotal as computed above, and* 0.5 *is the probability of guessing the "right" value in this case). So agent* 1 *has no incentive to compute.*

What the above example demonstrates is that a naive way of approaching agents might not work. Indeed, if all agents are approached simultaneously then we will face a similar problem. This motivates the careful discussion of sequential elicitation mechanisms, i.e. the construction of mechanisms that approach agents in a well designed sequence.

The design of mechanisms that deal with agent incentives is the subject of study of the theory of mechanism design in game theory. Most models discussed in that theory look at mechanisms that actually communicate with all agents simultaneously. Sequential mechanisms discussed in that literature are typically multi-stage games where the designer/center does not access agents sequentially, but the agents themselves may choose actions sequentially. In this paper we consider mechanisms that approach one agent after the other (we refer to these as sequential mechanisms).

The intuition behind the adoption of sequential mechanisms is as follows. Assume there are agents with low (access / computation / learning) costs. These agents can be approached first. It is possible that based on their replies the desired multi party computation can be carried out. However, if this is not the case then, intuitively, the impact of the other agents on the result of the computation (their pivotalness) increases, and so the incentive to incur the cost and compute increases. We demonstrate this argument in the following example:

**Example 2** *Consider* 4 *agents, each having a secret of either* 0 *or* 1, *drawn independently with equal probabilities. The agents would like to know whether they have a consensus or not. Assume three of them have very low computation costs (say, zero) and the fourth has a cost of* 0.4. *A mechanism that approaches the forth agent first (or all agents simultaneously) may fail to compute correctly as the fourth agent will choose to 'guess' his secret (by guessing his payoff is 0.875, whereas by accessing the secret the payoff is* 0.6). *However, if the mechanism approaches the first three before it approaches the* $4^{th}$ *agent, then we either may learn the true value (which is the case if there is lack of consensus already among these three) or may update the fourth agent about the previous replies. In this case the fourth agent's 'pivotalness' increases and so by guessing he expects a payoff of 0.5, compared to 0.6, when not guessing. Consequently, the fourth agent will access his secret and the mechanism will surely compute correctly.*

The class of multi party computations we study is that of anonymous functions. An anonymous function is one where the function's value does not depend on the identity of the agents but on the secrets only. In other words, a permutation of agents' secrets will not change the value of the function. This class of functions is quite elementary and often used in models. Among the anonymous functions are majority, consensus, unanimity, average, variance, order statistic, percentile and more.

The basic new terminology we introduce for sequential mechanisms is that of an 'appropriate mechanism'. A mechanism is deemed appropriate if for all realizations of secrets the function will be computed correctly, in an equilibrium (additionally, there may be equilibria of appropriate mechanism where this is not the case). In the paper we concentrate on mechanisms that are to act in fully revealing (e.g. broadcast) environment. In such setting the agent to be approached, and its response, is observed by all agents. We believe that this is a most appropriate setting for a basic model of information elicitation in multi-agent systems. We later discuss the effects that may be suggested by considering other models and extensions.

Informally, the main contributions we report on are as follows:

- We introduce a general setting for information elicitation in multi-agent settings, captured by the "multi-party computation game".

- We provide an efficient algorithm for constructing an appropriate mechanism, if such mechanism exists.

- We provide a characterization for the existence of appropriate mechanisms.

- We provide an efficient verification algorithm for the existence of an appropriate mechanism.

- We show that by allowing private communication, as well as by allowing probabilistic mechanisms, we can sometime improve the prospect of finding appropriate mechanisms.

## 2 Model

Let $N = \{1, \ldots, n\}$ be a finite set of agents. Each agent $j$ has a unique secret, $s_j \in \{0, 1\}$, that he may compute. Let $0.5 \leq q < 1$ be the prior probability of $s_j = 1$ and assume these events are independent.[3] Agents may compute their own secrets, however, computation is costly and agent $j$ pays $c_j \geq 0$ for computing $s_j$. Without loss of generality we shall assume $c_1 \leq c_2 \leq \ldots \leq c_n$ (in words, agents are ordered by their costs).

Agents are interested in computing some joint binary parameter (e.g., the majority vote or whether they have a consensus) that depends on the vector of private inputs. Let $G : \{0, 1\}^n \to \{0, 1\}$ denote the desired computation. Each agent $j$ has a utility of $v_j$ from learning the real value of $G$. We will assume that $v_j > c_j$, otherwise the agent faces no dilemma (we assume no side payments).

In the exposition we will use the convention that $v_j = 1$. This is done without loss of generality, as the more general case where $v_j > c_j > 0$, is equivalent to the case where the value of agent $j$ is 1 but the cost is $\frac{c_j}{v_j}$.

A central designer elicits the agents' secrets, computes $G$ and reports the computed value of $G$ back to each agent. In this setup each agent faces a dilemma of whether to compute his private secret $s_j$, at a cost of $c_j$, or perhaps to submit a guess to the central designer. The desired property of a mechanism is the correct computation of $G$, which is done through the elicitation of secrets from sufficiently many agents. One should note that as the cost of each agent's computation, $c_j$, is lower than the gain from computing $G$, the socially optimal outcome is to compute. However, free riding of agents may undermine the ability to reach the social optimum.

---

[3]The results apply for all $0 < q < 1$, however we assume w.l.o.g. that the probability of the event $\{s_j = 1\}$ is greater than or equal to that of $\{s_j = 0\}$.

In the introduction we considered an example where computation and truth revealing is not in equilibrium. The following example illustrates another situation.

**Example 3** - *Let $G$ be the parity function, $q = 0.5$, and $c_j = 0.4$ for all $j = 1, \ldots, 11$. Once again, consider a simple mechanism that asks the agents in arbitrary order for their secret and computes $G$. In this example each agent, when approached, is pivotal and therefore all agents computing is an equilibrium, and so $G$ will be computed correctly.*

### 2.1 Sequential Mechanisms

It is interesting to note that there may be a strict advantage in approaching agents sequentially. The intuition is that agents with a high computational cost may not be willing to compute, unless convinced they are pivotal. However, if some agents with low computation cost have already provided their secret the other agents may face one of two situations. Either $G$ can be computed from previous replies or it cannot. In the latter case the remaining agents may be more pivotal, perhaps sufficiently pivotal to justify a costly computation.

We now model mechanisms that approach agents sequentially. We will assume that the communication between the center and the agents is fully revealed by all agents. This can be associated with having broadcast communication. Moreover, we find this as an excellent model for introducing the idea of information elicitation in multi-agent systems, since it allows a simple setting where basic fundamental results can be obtained. Nevertheless, in Section 5 we discuss the possible advantages of having private channels.

A *sequential mechanism* (or mechanism) is an ordering of the set of agents, where the $k^{th}$ agent in the order is defined according to the reply of its predecessors. Furthermore the $k^{th}$ agent is provided with the replies of its predecessors.

Formally, let $H_i = \{0, 1\}^i$ be the set of histories of

length $i$, for $1 \leq i \leq n$, and $H_0 = \Lambda$ where $\Lambda$ is the empty (null) history. Let $H = \cup_{i=0}^{n} H_i$. A sequential mechanism is a pair $(g, f)$ where $g : H \to N$ determines the agent to be approached, and $f : H \to \{0, 1, *\}$ is a function that expresses a decision about whether to halt and output either 0 or 1, or continue the elicitation process (denoted by $*$). We will assume that if $g(h) = j$ then $g(h') \neq j$ for every $h'$ where $h$ is a prefix of $h'$, i.e. an agent is approached at most once.

The action space of each agent, $j$, is the set {compute, don't compute} $\times$ $\{0, 1\}$. The first coordinate refers to whether the agent chooses to go through the costly computation and the second coordinate is what the agent chooses to inform the central mechanism.

Note that this implies that each agent has 6 (and not 4) actions: Don't compute and report 0, Don't compute and report 1, Compute and report 0, Compute and report 1, Compute and report the true computed value, and Compute and report a false value. Let us denote by $\Gamma$ the set of actions.

A pure strategy for agent $j$, $x_j : 2^H \to \Gamma$, assigns an action to each possible subset of histories, and a (mixed) strategy, $X_j : 2^H \to \Delta(\Gamma)$, assigns a probability distribution over $\Gamma$. The parameter $q$, alongside the tuple of (mixed) strategies, $\{X_j\}_{j=1}^n$, determines the probability that $G$ will be computed.

An equilibrium for the mechanism $\mathcal{A} = (f, g)$, is a vector of $n$ strategies, one for each agent, such that each agent's strategy is the best response against the other agents' strategies.

We seek mechanisms which can compute the true value of $G$ in equilibrium. In fact, it is required that the mechanism computes $G$ with certainty. Therefore we seek mechanisms that induce sufficiently many agents to compute their true secret, in order for $G$ to be computed. Note that in many cases $G$ may be computed with partial information. For example, in the case of a consensus function it is sufficient to elicit information sequentially until we get 2 different replies, which are truthful.

**Definition 1** *A mechanism $\mathcal{A}$ is* appropriate *for $G$ at $0.5 \leq q < 1$ if there exists an equilibrium where $G$ can surely be computed for all vector of agents' secrets. Such an equilibrium is referred to as a* computing equilibrium, *and $\mathcal{A}$ is called q-appropriate.*

## 3 Constructing appropriate mechanisms

Having introduced the general setting, our aim is now to deal with the actual construction of appropriate mechanisms. We face two major challenges:

1. We wish to have an effective procedure for characterizing the behavior of an appropriate mechanism, if exists. In particular, we should come up with a technique for deciding on the "next agent to be approached" as a function of the history.

2. We wish to have an efficient procedure for checking the existence of an appropriate mechanism.

In this section we deal with the first issue, while the other one will be discussed in the next session. We now introduce the HCF (High Cost First) algorithm for (dynamically) ordering the agents. We omit some of the secondary details to be discussed in the full paper.

- Step One - For each possible prefix, i.e. string of 0 and 1's, of length in between 0 and $n-1$, compute the probability of being pivotal, conditional on agents being truthful. By knowing this probability we will have an upper bound on the cost of an agent that is expected to compute in equilibrium. Note that if $G$ is anonymous then checking this is polynomial, given a particular prefix.

- Step Two - Let us denote by $Z_j$ the event that $j$ is pivotal, and consider the following recursive structure. For any given set of agents and costs choose the agent to move first as follows - Consider all agents with a cost low enough to justify computing (namely all $j$ such that $1 - c_j \geq \text{Prob}(Z_j) \cdot q + P(Z_j^c) \cdot 1$) and approach the agent with the highest cost among these. Notice that this computation for an anonymous function is polynomial.

- Use this procedure to allocate the first agent ($\sigma_1$). Depending on the reply of $\sigma_1$ you end up with one of two trees. Apply the same procedure again to each tree, where at each time you are not allowed to allocate agents that have been already allocated, and so on and so forth.

Using HCF we can show:

**Theorem 1** *Let $G$ be an anonymous function. Assume it is common knowledge that there exists an appropriate mechanism for $G$, then the mechanism induced by the HCF algorithm has a computing equilibrium.*

**Proof (sktech):** We use an inductive argument, on the total number of agents. For $n = 1$ this is straightforward. Assume the statement of the theorem holds for all problems with $n = N$ agents and consider a problem with $n = N + 1$ agents.

It is common knowledge that the original problem has an appropriate mechanism. Assume agent $k$ is the first agent in that appropriate mechanism. We now know two things. First, that $1 - c_k \geq \text{Prob}(Z_k) \cdot q + P(Z_k^c) \cdot 1$, and second, that in the two problems induced, following $k$'s reply, there are $n = N$ agents (all $N + 1$ agents, but $k$) and it is common knowledge that an appropriate mechanism exists.

Consider agent $\sigma_1$ that was chosen by our algorithm. By definition $c_{\sigma_1} \geq c_k$ (note that due to anonymity $\text{Prob}(Z_{\sigma_1}) \cdot q + \text{Prob}(Z_{\sigma_1}^c) \cdot 1 = \text{Prob}(Z_k) \cdot q + \text{Prob}(Z_k^c) \cdot 1$). Hence, if agent $\sigma_1$ moves first the existence of an appropriate mechanism for the two problems it induces is common knowledge. Notice that using $c_{\sigma_1}$ instead of $c_k$ can be only helpful, and each sub-case will remain appropriate.

Now, by our induction hypothesis, the algorithm induces an appropriate mechanism for each of the two games induced by $\sigma_1$'s reply, and so a computing equilibrium exists. It is therefore a best reply for agent $\sigma_1$ to compute as well.　　QED

Note that the algorithm suggested can be implemented on-line. In other words, computation may take place only along the realized path. This is particularly interesting from a complexity point of view, as the number of histories along a particular path is $n$, whereas the total number of histories in $H$ is of the order of magnitude of $2^n$ (and a naive off-line algorithm will have to refer to each of them).

The above result hinges on the existence of an appropriate mechanism. However, verifying whether such a mechanism exists may be a non trivial task. One intuitive way to proceed with such a verification process is to apply the algorithm suggested in the proof of Theorem 1 and eventually check whether the resulting mechanism is appropriate. However, this naive approach turns out to be exponentially complex, as each play path along the tree will have to be checked. The next section discusses the complexity of algorithms for verifying the existence of appropriate mechanisms.

## 4  The existence of appropriate mechanisms

In this section we discuss what we believe to be a central technical problem in information elicitation in multi-agent systems. We wish to characterize conditions for the existence of computing equilibrium, i.e. find necessary and sufficient conditions for the existence of an appropriate mechanism. Moreover, we are interested in making an algorithmic use of this characterization. Indeed, we will show such characterization that leads to a polynomial algorithm for checking whether an appropriate mechanism exists.

We are interested in an efficient algorithm that, given the values $c_1, \ldots, c_n$, $q$, and a succinct representation of an anonymous function (namely the number of 1's in the input needed for having the value 1. Note that this is a list of up to $n + 1$ numbers), decides whether there exists an appropriate mechanism or not. The algorithm we propose is composed of a "verification criterion" and a low complexity algorithm verifying whether this criterion is satisfied. We begin by treating the criterion issue, which is based on the appropriateness of the HCF algorithm. Later we turn to translate this into an efficient verification algorithm.

Let $G$ be an arbitrary anonymous function. Consider a directed graph $G' = (V, E)$ where the set of nodes is $V = \{(i, k) : i, k \in Z_+, 0 \leq k \leq i \leq n\}$, and the set of edges is $E = \{((i, k), (i + 1, k)) : 0 \leq i \leq n - 1, k \leq i\} \cup \{((i, k), (i + 1, k + 1)) : 0 \leq i \leq n - 1, k \leq i\}$. Intuitively, this graph describes the possible states of information, when approaching agents while computing the value of an anonymous function, and the possible state transitions. Every state describes how many agents have been approached and how many 1's have been heard so far. In particular, the node $v = (j, l)$ in $G'$ is interpreted as the event that $j$ agents have been approached and $l$ 1's (and $j - l$ 0's) have been reported.

For each such node we may ask whether the value of $G$ can be computed, or not. Let $\bar{G} = (\bar{V}, \bar{E})$ be the graph induced by reducing $G'$ to include only nodes where the value of the function $G$ is still unknown. Notice that $\bar{G}$ is a DAG.

For every node in $\bar{G}$, $v$, let $c(v)$ be the largest $i$, for which $1 - c_i \geq \text{Prob}(Z_i|v) \cdot q + P(Z_i^c|v)$, where $Z_i$ is the event that agent $i$ is pivotal for the function $G$. Namely, if $c(v) = i$ then every agent $j \leq i$ will satisfy the above inequality (i.e., will have incentive to check its secret, assuming the rest of the agents check their secrets, when approached), and every agent $j > i$ will not satisfy the above inequality (i.e. will not have incentive to check its secret, assuming the rest of the agents check their secrets). Assume that for some node $v$ in $\bar{G}$, $c(v)$ is undefined, then this implies that no agent will have incentive to check and so an appropriate mechanism will not exist. Therefore, we will

consider below only cases where $c(v)$ can be defined for every node $v$.

Our algorithm is based on the following criterion:

**Theorem 2** *Let $\bar{G} = (\bar{V}, \bar{E})$ be the directed graph associated with an anonymous function $G$, $0.5 \leq q < 1$, and let $c(v)$, for any $v \in \bar{V}$, be defined as above. Then, there exists a q-appropriate mechanism iff there does not exist a path in $\bar{G}$ originating from $(0,0)$ and leading to some end-node $v$, $(v_1 = (0,0), v_2, \ldots, v_l = v)$, for which there exists some $1 \leq i \leq n$ such that $|\{v_j : c(v_j) \leq i\}| > i$.*

**Proof (sketch):**

Proof of the first direction - Assume there exists a path in $\bar{G}$ originating in node $(0,0)$, and leading to some end node $v$, $(v_1 = (0,0), \ldots, v_l = v)$, and some $i$ such that $|\{v_j : c(v_j) \leq i\}| > i$ then since $v$ is a reachable state (if $v = (k,l)$ then we will reach $v$ whenever after approaching $k$ agents we have been reported on $l$ 1's and $k - l$ 0's) the above condition immediately implies that there will be an agent who will not check its secret while this agent is pivotal (since the graph consists only of nodes where the value of $G$ is still undetermined). This means that in this case no sequential mechanism is appropriate.

Proof of the second direction - Assume an appropriate mechanism does not exist. Then, the HCF mechanism will not be appropriate. Let $v$ be a node where the HCF mechanism halts and let $(v_1 = (0,0), \ldots, v_l = v)$ be the specific path leading to $v$. As HCF halts we conclude that it has already placed all agents $1, 2, \ldots, c(v)$ along the path $(v_1 = (0,0), \ldots, v_l = v)$. Denote by $i$ the maximal number for which all agents $1, 2, \ldots, i$ have been placed by the HCF. In particular this means that $i+1$ was not placed. Obviously $i \geq c(v)$. For any $j \leq i$ let $v_j$ be the node at which $j$ was placed.

We claim that $c(v_j) \leq i$. Assume this is incorrect and $c(v_j) > i$. In this case HCF would have chosen $i+1$, which is available at this node, or a higher cost agent, and not agent $j$ (recall $j \leq i$). Therefore along the path $(v_1 = (0,0), \ldots, v_l = v)$, $|\{v_j : c(v_j) \leq i\}| > i$, and so any path from the origin to an end node which prefix is $(v_1 = (0,0), \ldots, v_l = v)$ satisfies the property that there exists some $1 \leq i \leq n$ such that $|\{v_j : c(v_j) \leq i\}| > i$. QED

We use the appropriateness criterion of Theorem 2 to present an efficient algorithm for verifying appropriateness:

**Theorem 3** *Let $G$ be an anonymous function, and consider a set of $n$ agents, with costs $c_1, \ldots, c_n$, and a parameter $0 < q < 1$, then there exists a polynomial algorithm for checking the existence of an appropriate mechanism*

**Proof (sktech):**

1. We construct the graph $\bar{G}$ and compute the $c(v)$'s. Notice that since the function is anonymous, these computations are polynomial. If $c(v)$ can not be determined for some node then there is no appropriate mechanism.

2. For any end-node $v = (k, l) \in \bar{V}$ we check whether there is a path of length $j+1$ annotated by $c(v)$'s of value $j$ or lower connecting $(0,0)$ to that node (for any $j$ between 1 and $n$)
   - Construct a graph where every node $v'$ which is associated with $c(v') \leq j$ gets the weight 1, and each other node gets the weight 0.
   - Look for the maximal weighted path (i.e. a path where the sum of weights is maximized) between $(0,0)$ and $v$ and check if it is of value $j+1$ or more. Notice that this is polynomial.

3. If no such path is found (for any node $v$) then there exists an appropriate mechanism, and otherwise there is no appropriate mechanism.

The proof follows from Theorem 2, the feasibility of generating the graph $\bar{G}$ and the efficiency of standard path finding algorithm. Details omitted. QED

Notice that theorems 2 and 3 introduce an interesting complexity-theoretic lesson. These theorems do not tell us that there exists an efficient procedure that outputs an appropriate mechanism. Indeed, the size of the description of such a mechanism may be exponentially large. However, these theorems give us an efficient procedure for studying the *existence* of such a mechanism, although the mechanism, if exists, may not be off-line (efficiently) constructed. In addition, given the results of section 3, *if* our (efficient) procedure verifies that an appropriate mechanism exists, then it can be (efficiently) constructed on-line (in a step by step fashion). We believe that this situation, where the decision problem and the search problem have such inter-play, sheds an interesting light on the study of the complexity of mechanisms for information elicitation.

## 5 Extensions: Privacy and Randomness

Our model introduces information elicitation in the framework of a basic setting, where each agent is pro-vided with full information about the history until it has been approached. While this is a natural basic model, one may consider extended models, where agents may be provided only with partial information. We now briefly discuss this extension.

### 5.1 Privacy: Providing Partial Information

Partial information is modelled by having mechanisms where the information provided is also a function of the history. Formally, this can me modelled by adding to the definition of the mechanism another function: $h : H \to 2^H$ where this function states which information will be provided as a function of the history so far. Notice that in general the mechanism may "cheat" about the history observed so far; an agent's state of information will be determined by his knowledge of the mechanism and the information provided to him by the mechanism (the formal definition is therefore as in the theory of knowledge, as used in AI and distributed computing (Fagin *et al.* 1995)). Notice that this general definition enables to express the simultaneous mechanism (by always providing all agents with the information $2^H$). Moreover, it allows to approach agent $i$ after two different histories $(k, l)$ and $(k', l')$ without the agent knowing which history is the actual one (as long as one of these histories is not a prefix of the other). We can then define computing equilibrium and appropriate mechanisms as before. We can show:

**Proposition 1** *There exist anonymous functions, $c_i$'s and $q$, for which there does not exist an appropriate mechanism in the full information setting, but there exists an appropriate mechanism in the partial information setting.*

### 5.2 Probabilistic Mechanisms

Another question of interest relates to a random/probabilistic behavior of the information elicitation mechanism. In particular, we ask whether probabilistic mechanisms may be helpful in leading to computing equilibrium. This topic is not an issue when considering our basic ("broadcast") setting; in this case it is easy to see that since the center and the agents have the same information about the history, probabilistic decisions (about the agent to be approached) are not helpful. However, probabilistic mechanisms may be helpful when considering the partial information setting. We can show:

**Proposition 2** *There exists an anonymous function, $c_i$'s and $q$, for which there does not exist an appropriate deterministic mechanism in the partial information setting, but there exists an appropriate probabilistic mechanism. The result remains valid also when we require all $c_i$'s to be identical.*

Notice that this result shows the power of sequential mechanisms also for the case where all agents have identical costs. Indeed, it can be shown that when all costs are equal then any deterministic sequential mechanism can not do better than the simultaneous simple mechanism. However, Proposition 2 shows that in the case of equal costs a probabilistic sequential mechanism can be more effective than such deterministic mechanisms. This serves as further evidence to the power of information elicitation using sequential mechanisms.

## 6 Discussion

This paper introduced sequential information elicitation in multi-agent systems, using the multi-party computation game. The novelty of our approach stems from treating multi-party computation as a public good setting, where costly computation of private inputs leads to free-riding problem, and from the introduction and study of mechanisms (and in particular sequential mechanisms) for overcoming this problem.

Our paper emphasizes the existence of appropriate mechanisms. In particular we show an algorithm for verifying the existence of an appropriate mechanism, as well as provide a criterion for existence. Most of this is carried out for the case of broadcast communication channels. The issue of private channels, which is discussed only briefly, is a subject of a complementary study (Smorodinsky & Tennenholtz 2003). Although the discussion of private channels can be viewed as an extension of the work on broadcast communication, the previously mentioned work does not deal with the problem of existence. While for the question of existence we provide in this paper a general (and positive, from the computational perspective) answer, the study of random mechanisms and their power, introduced in the previous section, is a subject that deserves much further attention.

Multi-party computation is a central topic in computer science. The main objective discussed in the literature on multi-party computation is to devise protocols for collective computation of a function's value without having any information revealed to the parties, beyond the function's value (see Goldreich (Goldreich 1998) for a recent general overview, and Linial (Linial 1994) for a discussion of a game-theoretic perspective on that issue). Although game-theoretic in nature, multi-party computation does not include standard game-theoretic analysis. A complementary perspec-tive, which does adopt a game-theoretic approach to multi-party computation, has been recently introduced by Shoham and Tennenholtz (Shoham & Tennenholtz 2002). In their setting, titled non-cooperative computing, an agent's utility is effected by two factors: a primary objective of computing the function, and a secondary objective of preventing others from computing it. They provide full characterization of the boolean functions that can be non-cooperatively computed under various assumptions on the economic setting (e.g. private values vs. correlated values) and the algorithmic setting (e.g. deterministic vs. probabilistic algorithms). Most recently, work that attempts to combine secure multi-party computation and non-cooperative computing has been introduced by McGrew, Porter and Shoham (McGrew, Porter, & Shoham 2003), and work that deals with non-cooperative computing where there is no center has been introduced by Halpern and Teague (Halpern & Teague 2004). This work can be viewed as dealing with externalities in multi-party computation, and does not deal with the free-riding problem and the role of sequential information elicitation.

Recent work in AI has been concerned with algorithms for preference elicitation (see e.g. (Conen & Sandholm 2001; Boutilier *et al.* 2003)). Our work contributes also to the literature on techniques for preference and information elicitation. However, although some of that work adopts a game-theoretic perspective (see e.g. (Shoham & Tennenholtz 2001)), it does not deal with the fact private information might be costly to acquire, and with the fundamental free riding problem this issue introduces in a strategic multi-agent setting.

The study of free riding has a long tradition in Economics and Game Theory, in particular in the context of the Public Good problem (e.g., chapter 13 in Mas-Colell, Whinston and Green (Mas-Colell, Whinston, & Green 1995)). In many models it is shown that free riding is sufficiently destructive to prevent socially optimal outcome (e.g., Rob (Rob 1989) and Mailath and Postlewaite (Mailath & A. 2000)). In this paper, we take a more constructive approach and seek mechanisms that overcome the free riding problem and result in the efficient outcome, which is the correct execution of the multi-party computation.

In future work we plan to make the setting more tightly related to preference elicitation. One way to obtain this is by considering situations where the agent's private information captures also its preference on the function's outcome. It may be also interesting to consider minimizing the cost of elicitation. Other cases of interest include handling less symmetric cases, e.g. when agents may have different probability distributions on possible secrets.